\documentclass[useAMS,usenatbib]{mn2e}
\usepackage{graphicx}
\usepackage{txfonts}
\usepackage{longtable,lscape}
\newcommand{\HeII}{He\,{\sc ii}}

\newcommand{\HII}{H\,{\sc ii}}

\newcommand{\SII}{[S\,{\sc ii}]}

\def\p0{\phantom{0}}

\def\arcsec{\hbox{$^{\prime\prime}$}}

\newcommand{\OI}{[O\,{\sc i}]}

\newcommand{\OIII}{[O\,{\sc iii}]}
\newcommand{\NII}{[N\,{\sc ii}]}

\begin{document}
 \title{The Physical Parameters of the Micro-quasar S26 in the Sculptor Group Galaxy NGC~7793}

 %  \subtitle{ }

   \author[Dopita et al.]{M. A. Dopita$^{1,2}$, J. L. Payne$^{3}$, M. D. Filipovi\'c$^{3}$, \&  T. G. Pannuti$^{4}$\\
   michael.dopita@anu.edu.au\\
   $^{1}$Research School of Astronomy and Astrophysics, Australian National University,
              Cotter Road, Weston, ACT 2611, Australia\\
  $^{2}$Astronomy Department, King Abdulaziz University, P.O. Box 80203, Jeddah, Saudi Arabia\\
  $^{3}$University of Western Sydney, Locked Bag 1797, Penrith South DC, NSW 2751, Australia\\
  $^{4}$Department of Earth and Space Sciences, Space Science Center, Morehead State University, Morehead, KY 40351, USA\\
          }

\date{Accepted . Received }

\pagerange{\pageref{firstpage}--\pageref{lastpage}} \pubyear{}

\maketitle

\label{firstpage}

\begin{abstract}
NGC 7793 - S26 is an extended source (350 pc $\times$ 185 pc) previously studied in the radio, optical and \mbox{x-ray} domains. It has been identified as a micro-quasar which has inflated a super bubble. We used Integral Field Spectra from the Wide Field Spectrograph on the ANU 2.3 m telescope  to analyse spectra  between 3600--7000 \AA.  This allowed us to derive fluxes and line ratios for selected nebular lines. Applying radiative shock model diagnostics, we estimate shock velocities, densities, radiative ages and pressures across the object. We show that S26 is just entering its radiative phase, and that the northern and western regions are dominated by partially-radiative shocks due to a lower density ISM in these directions. We determine a velocity of expansion along the jet of 330~km~s$^{-1}$, and a velocity of expansion of the bubble in the minor axis direction of 132~km~s$^{-1}$. We determine the age of the structure to be $4.1\times10^5$~yr, and the jet energy flux to be  $ (4-10)\times10^{40}$~erg s$^{-1}$ The jet appears to be collimated within $\sim0.25$~deg, and to undergo very little precession. If the relativistic $\beta \sim 1/3$, then some 4~M$_{\odot}$ of relativistic matter has already been processed through the jet. We conclude that the central object in S26 is probably a  Black Hole with a mass typical of the ultra-luminous X-ray source population which is currently consuming a fairly massive companion through Roche Lobe accretion.
 \end{abstract}
 
\begin{keywords}
black hole physics --  shock waves -- galaxies: individual(NGC7793) -- stars: winds, outflows -ISM: abundances, bubbles, jets and outflows  -- X-rays: binaries         
\end{keywords}                  
%________________________________________________________________

\section{Introduction}

The Sculptor Group galaxy NGC 7793 is a  SA(s)d  spiral galaxy located at RA (J2000) 23h57m49.83s,Dec.  --32d35m27.7s. It has a distance estimated from the tip of its Red-Giant Branch of 3.6~Mpc (Jacobs et al. \citeyear{Jacobs09}) or 3.44~Mpc estimated from its Cepheid stars (Pietrzynski \citeyear{Pietrzynski10}). The foreground Galactic Extinction E(B--V) is estimated at 0.019 magnitude (Schlegel et al. \citeyear{schlegel98}). In this paper, we adopt a distance of 3.5~Mpc.

The elongated \HII\ region within it, N7793-S26, was discovered by Blair \& Long (\citeyear{blairlong97}) using interference filters isolating the H$\alpha$ + \NII\ lines and the \SII\ lines in order to   identify Supernova Remnants (SNRs).  They described the object as a extended oval emission region with a  high \SII\ to H$\alpha$ ratio and a size of 185 $\times$ 350 pc (based on the distance of 3.5~Mpc rather than the 3.38 Mpc used in their paper). The major axis of this nebula is oriented southeast to northwest direction. They noted there was no evidence of an interior  star cluster, but they suggested that multiple supernovae (SNe) may be in the process of creating a super bubble, which would account for its unusually large size.
    
  Read \& Pietsch (\citeyear{readpietsch99}) reported \mbox{X-ray} observations of P8 coincident within 3\arcsec\ to N7793-S26, best fitted by  a low-temperature thermal model.  They could not rule out  the possibly of time variability as seen in their Fig. 4   {\it ROSAT}  light curves.  Very Large Array (VLA) radio observations at 6 and 20 cm followed, and were reported in Pannuti et al. (\citeyear{pannuti02}).  They noted the object is composed of both a bright point and extended source, designated together as S26.  Their discussion of the object  is quite extensive, noting it to be the only evolved stellar  candidate in N7793 detected in the \mbox{X-ray}, optical and radio domains. The object was seen in Chandra and spatially resolved as a triple source for the first time by Pakull \& Gris\'e (\citeyear{pakull08}), and was further studied by Pakull et al. (\citeyear{pakull10}) and Soria et al. (\citeyear{soria10}). Later, Pannuti et al. (\citeyear{pannuti11}) also detected the X-ray counterpart to S26 in their analysis of the Chandra data obtained in NGC7793.
   
Pakull \& Gris\'e (\citeyear{pakull08}) noted that a significant fraction of Ultra-Luminous X-ray sources appear to be embedded in extended ($\sim 100$pc) bubbles of shocked gas, and they identified N7793-S26 as belonging to this class of objects. Pakull et al. (\citeyear{pakull10}) further suggested that N7793-S26 is powered by black hole with a pair of jets, similar to SS433; but twice as large and more powerful.  They suggest S26 has a structure similar to a Fanaroff-Riley type II active galaxy with a \mbox{X-ray} and optical core,  \mbox{X-ray} hot spots and radio lobes encased in an optical  and   \mbox{X-ray} cocoon. The core can be fit by a power law in the  0.3--10 keV energy band, and has a luminosity of $L_{0.3-10} \approx 7 \times 10^{36}$ ergs s$^{-1}$. There are two X-ray bright hot spots on either side of the central source, which appear to be thermal in nature with a combined luminosity of $L_{0.3-10} \approx 1.8 \times 10^{37}$ ergs s$^{-1}$.  Using \HeII\ narrow-band observations, they find significant emission at 4686 \AA\ (He$^{2+} \rightarrow$ He$^{1+}$) spatially distributed in a similar manner to H$\alpha$.  A narrow slit spectrum taken near the core suggested an expanding shell with a maximum expansion velocity of $\sim250$ km s$^{-1}$.
   
 Soria et al. (\citeyear{soria10}) used the Australia Telescope Compact Array (AT) to resolve the radio lobe structure and to map the spectral index in the  radio cocoon of S26.  The radio structure is strongly aligned to the axis defined by the three \mbox{X-ray} hotspots  While the radio spectral index in the NW and SE lobes is approximately --0.7 to --0.6, it is flatter (--0.4 to 0.0) across most of the cocoon and becomes inverted at the base of the jets near the core (0.0 to +0.4)\footnote{Spectral index here is defined as $\alpha$,  for $S_{\nu} \propto \nu^{\alpha}$ where $S_{\nu}$ is flux density and $\nu$, frequency.} suggesting a self-absorbed spectrum due to either free-free or Compton self-absorption.  The  total thermal energy  in the bubble is estimated to be $\sim 10^{53}$ erg.
   
In this paper, we present the results of integral field spectroscopy in the wavelength region 3600-7200\AA\, using the Wide Field Spectrograph (WiFeS)  (Dopita et al. \citeyear{Dopita07}, \citeyear{Dopita10}) mounted on the 2.3m telescope located at the Siding Spring Observatory. This allows the shock velocity, pre-shock density, ram pressure and radiative age to be derived across the face of the nebula, in turn permitting a much more accurate determination of the fundamental properties of the micro-quasar which is driving these shocks.

   \begin{figure*}
    \centering
     \includegraphics[width=0.90\hsize]{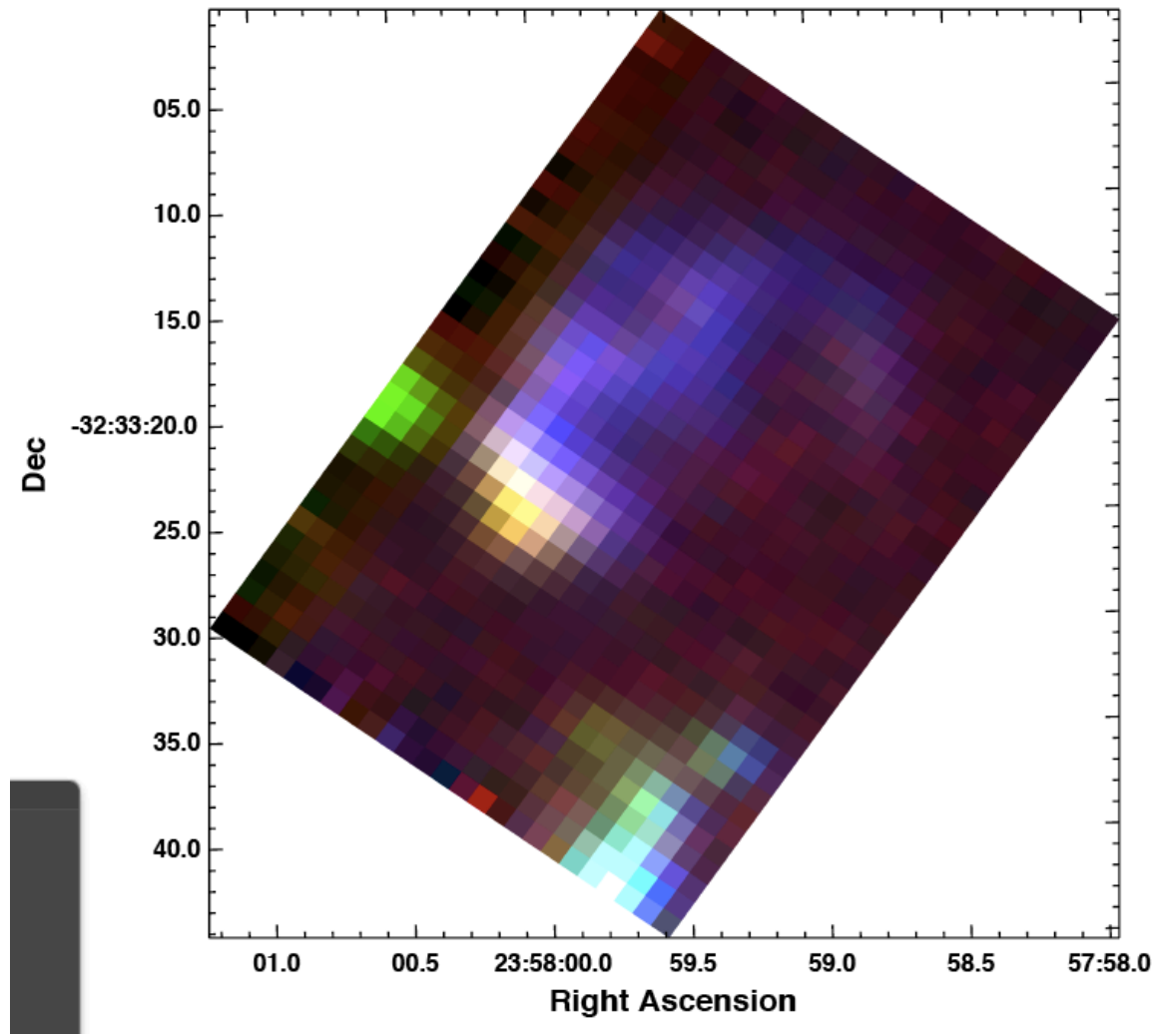}
    \caption{RGB (\SII, H$\alpha$, \OIII\ $\lambda 5007$) false color image of NGC7793-S26.  The y-axis has been rebinned to a final resolution of $1 \times 1$\arcsec, and the coordinates given are for Epoch J2000.  The purpose of this image is to  map these three emission regions for visual inspection. However, the strength of the \SII\ and H$\alpha$ lines in the SE jet region is apparent, as is the relative strength of the  \OIII\  lines in the fainter parts of the shell. The regions used for analysis are shown in Figure \ref{FigvoronoiS26}, and quantitive values for each of these emission lines in these regions can be found in Table \ref{Table1}. }
    \label{Fig1}%
    \end{figure*}

\section{Observations and Analysis}
Integral Field Spectroscopy (IFS) on N7793-S26 was performed on 13 December 2009 at Siding Springs Observatory using the 2.3 m Advanced Technology Telescope and its Wide Field Spectrograph (WiFeS). WiFeS provides a 25\arcsec\ $\times$ 38\arcsec\ field with 0.5\arcsec\ spatial sampling along each of 25\arcsec\ $\times$ 1\arcsec\ slits. The output format matches  the 4096 $\times$ 4096 pixel CCD detectors in both the red-side and blue-side cameras, and the instrument sensitivity is optimized for the blue and red sides of the
spectrum reaching a throughput (top of atmosphere to back of camera) of 35\%. We used the RT560 beam
splitter. The blue spectral range covers 3200--5900\AA, at a spectral resolution of $R \sim 3000$ ($\approx 100$ km s$^{-1}$).  In the red, the range is 5300--9800\AA,  and the resolution is $R \sim 7000$ ($\approx 45$ km s$^{-1}$).

Three 1000 second  exposures of S26 were taken under photometric conditions at PA (measured east of north) 145 degrees. The estimated seeing was excellent, $\approx1.3$\arcsec. Standard and telluric stars, flat field, bias and wavelength calibration data frames were also obtained on the same evening.

The data was reduced using the WiFeS data reduction pipeline based on NOAO (National Optical Astronomy Observatory)
IRAF software. This data reduction package was developed from the Gemini IRAF package (McGregor et al., \citeyear{mcgregor03}). Use of the pipeline consists of four primary tasks: {\sc wifes} to set environment parameters, {\sc wftable} to convert single extension FITS file formats to Multi-Extension FITS ones and create file lists used by subsequent steps, {\sc wfcal} to process calibration frames including bias, flat-field, arc and ÒwireÓ; and {\sc wfreduce} to apply calibration files and create data cubes for analysis.  We used the standard  star HD 26169  for flux calibration and the bright B-type star HIP 8352 for any necessary telluric corrections.

In Fig. \ref{Fig1} we show a false color RGB image of N7793-S26 in its proper orientation to the sky.  There is intense \SII\ 
emission in the southwestern hotspot, associated with the termination shock of the jet from the central object as seen in X-rays (Pakull et al. \citeyear{pakull10}). The elliptical shell of the bubble is, by contrast, much brighter in \OIII\ and the \OIII/H$\beta$ ratio increases as the Balmer line flux decreases. The shell brightness decreases going north on the eastern side, and is incomplete
on its western side. This, together with the asymmetry in the distance of the NW X-ray hotspot compared to the SW hotspot from the central X-ray source strongly suggest a systematic gradient in the density of the ISM, highest towards the SW side, and lower towards the NE.  This is reminiscent of the gradient seen across the system SS433 - W50 in the Galaxy  \citep{Zealey80}.

The other emission regions appearing in this figure (the green region close to 23:58:00.5 -32:33:18 (J2000) and the turquoise regions in the general vicinity of 23:57:59.7 -32:40:00 (J2000) are \HII\ regions which are unassociated with S26. These are useful in constraining the chemical abundances in the shock modelling described in the next section.

  \begin{figure*}
    \centering
    \includegraphics[width=0.90\hsize]{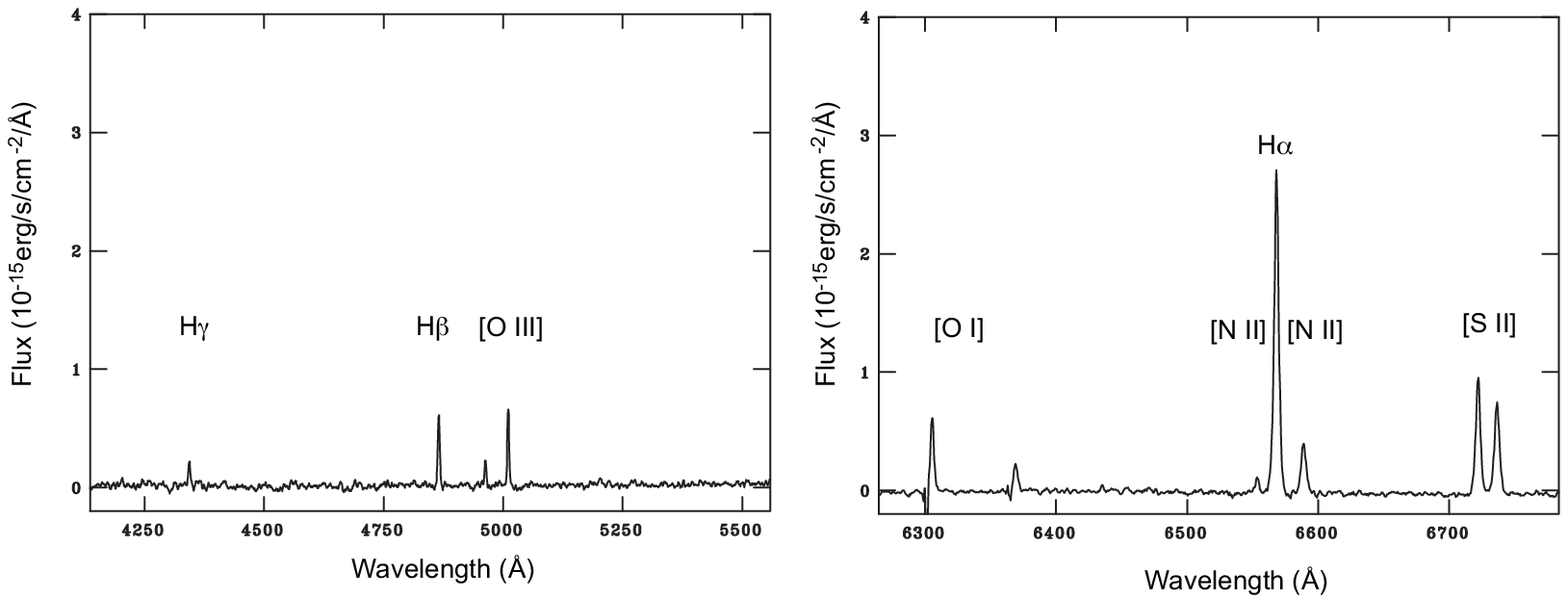}
    \caption{One-dimensional spectra of brightest $2\arcsec \times 2$\arcsec region in N7793-S26.  Note the large velocity dispersion in 	the lines. Note also that the N7793-S26 emission lines are redshifted ($\sim 228$ km s$^{-1}$) consistent with the reported 
	redshift of N7793 ($227$ km s$^{-1}$; Koribalski et al., 2004). }
    \label{spectrumfig}%
    \end{figure*}

The spectrum of the most intense $2\arcsec \times 2$\arcsec\ region is shown in Fig. \ref{spectrumfig}.  A classic shock-excited spectrum is revealed with both strong \OI\ and \SII\ emission relative to H$\alpha$ and velocity-broadened line profiles in all of the emission lines. This spectrum can be compared with that of the emission region in SS443 (\cite{Zealey80}; fig 2), in which the velocity dispersion is much lower, but the forbidden lines are much stronger with respect to H$\alpha$. The difference can be ascribed to two factors, first a much lower characteristic shock velocity in SS433 and second, lower chemical abundances in the ISM in S26.

\begin{figure*}
    \centering
    \includegraphics[width=0.80\hsize]{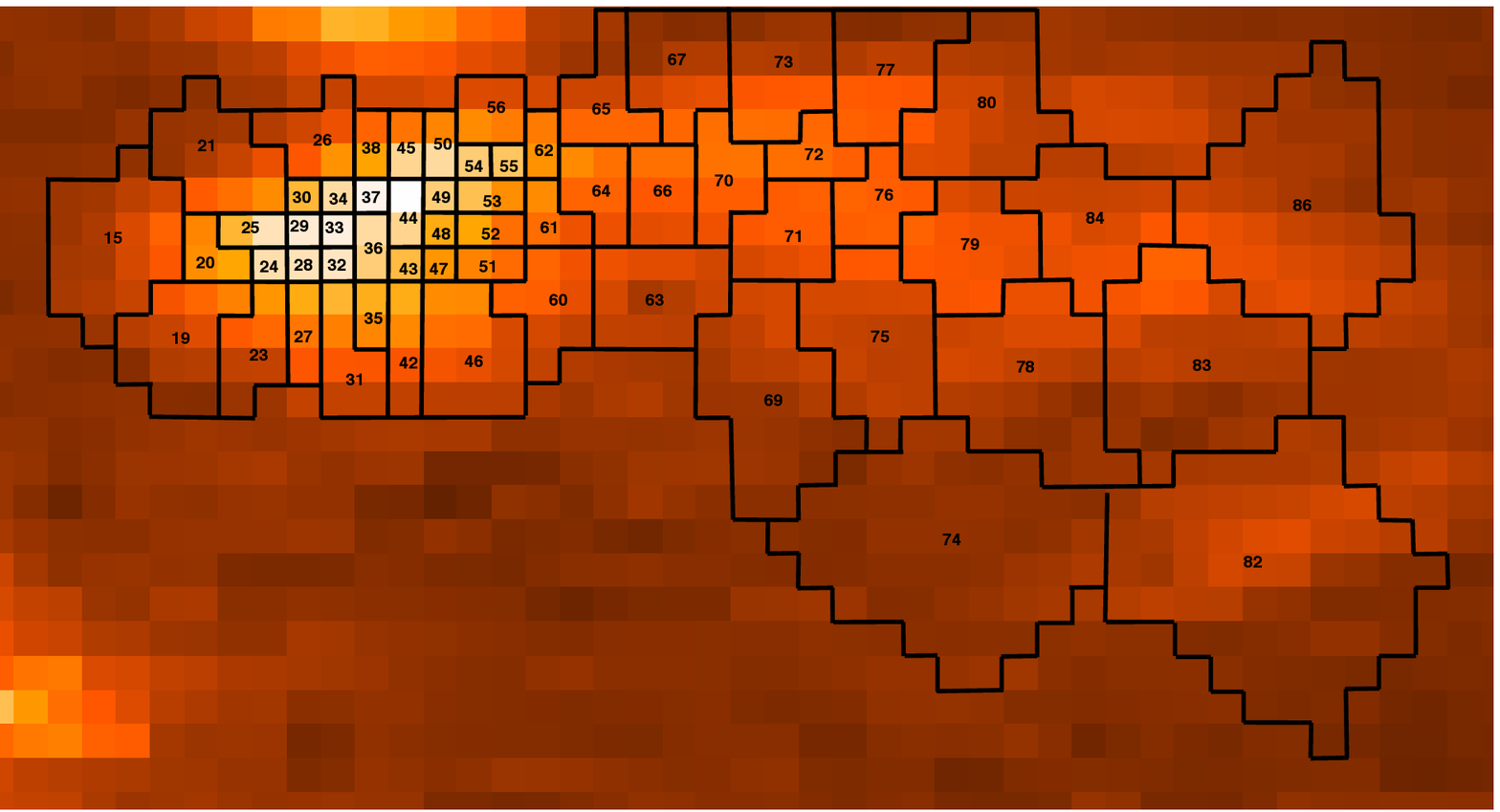}
    \caption{A portion of the elected Voronoi regions used for sampling overlaid onto part of the WiFeS H$\alpha$ image. The $0.5\times1.0$ arc~sec. pixels of the WiFeS instrument have ``stretched" this image in the horizontal direction compared with Figure 1. The regions 5, 6, 10, 12 and 13 are a complex of \HII\ regions, unassociated with S26. Apart from region 13 these are mostly located off  the edge of this image in the lower left region. These \HII\ regions were used to calibrate the chemical abundances appropriate to the analysis of S26.
                 }
    \label{FigvoronoiS26}
    \end{figure*}

Using QFitsView\footnote{Written by Thomas Ott and freely available at:\\www.mpe.mpg.de/$\sim$ott/dpuser/index.html} we were able to create a Voronoi Tessellation diagram based on a 30:1 H$\alpha$ signal to noise (S/N) ratio using the red data cube as shown in Fig. \ref{FigvoronoiS26}.  This was later used as a template to extract one dimensional spectra from NGC 7793-S26. After extraction, the IRAF task {\sc splot} allowed the determination of emission line flux and the velocity width based on Gaussian fits to the H$\alpha$ line profile.  Photometric measurement errors were calculated based on the root-mean-square noise estimated for each spectra; varying between an average of 2 percent for the stronger $\lambda 6563$ line and 23 percent for the weaker $\lambda 6364$ line.

In Table \ref{Table1}, we present flux densities (in units of $ 10^{-15}$ ergs cm$^{-2}$ s$^{-1}$) for selected emission lines corresponding  to the Voronoi extraction by number listed in Fig. \ref{FigvoronoiS26}.  The lines are labeled at rest wavelength and   include H$\beta$ ($\lambda 4861$), \OIII\ ($\lambda\lambda 4959, 5007$), \OI\ ($\lambda\lambda 6300, 6364$), \NII\ ($\lambda\lambda 6550, 6585$), H$\alpha$ ($\lambda 6563$) and \SII\ ($\lambda\lambda 3716, 3731$). Also included are log ratios \OIII$\lambda 5007$ / H$\beta$, \NII$\lambda 6585$ /  H$\alpha$, \SII$\lambda\lambda 6716, 6731$ / H$\alpha$ and 
H$\alpha$ surface brightness (ergs s$^{-1}$ cm$^{-2}$ sr$^{-1}$).

\section{Shock Analysis of S26}
Although the SE region of S26 shows a very normal shock-excited spectrum, the relatively strong \OIII\ emission in the northern part of the shell is harder to explain. There are two possible explanations -- either that this region is ionised by X-rays emanating from the central source, or that these regions contains only partially-radiative shocks which enhance the relative strength of the \OIII\ lines by cutting off the emission that would arise from the recombination zone of the shock. Such shocks were first considered by Dopita (\citeyear{dopita83}) in the context of the supernova remnants IC~443 and RCW~86, and the general behaviour of the line ratios of these ``truncated'' shocks were given. Later, Raymond et al.(\citeyear{Raymond88}) made a more complete study, and investigated the effects of density and shock velocity.

The possibility that the nebula is predominantly X-ray ionized can be readily eliminated on energetic grounds. The integrated luminosity of the X-ray sources in S26 is no greater than about $3\times10^{37}$erg s$^{-1}$ from observations taken with the Chandra satellite (Pakull, Soria \& Motch \citeyear{pakull10}). From the same reference, the H$\beta$ luminosity of S26 is given as $\sim 1 \times 10^{38}$erg s$^{-1}$. For comparison, the sum of the H$\beta$ flux in all the regions shown on Figure \ref{FigvoronoiS26} is $5.5\times 10^{37}$erg  s$^{-1}$. However, we are probably missing some flux, so we adopt the (Pakull, Soria \& Motch \citeyear{pakull10}) value here. Models for photoionised nebulae with an abundance of  $Z = 0.5Z_{\odot}$, appropriate to this region of NCG~7793, emit about 50 times the flux observed at H$\beta$ (L\'opez-S\'anchez et al. \citeyear{Lopez-Sanchez12}). Therefore, the total nebular luminosity has to be of order $5\times10^{39}$erg s$^{-1}$ -- over 100 times the observed X-ray luminosity. 

Apart from these energetic constraints, if X-rays were dominating the excitation, the strength of the \OI\ lines should be much greater in the SE portion of the nebula, since X-ray excited nebulae are characterised by extended warm, partially ionised tails in their \HII\ regions. This is expected on the basis of theoretical models as described in  Dopita \& Sutherland (\citeyear{dopita03}), and a fine example of this phenomenon was observed by Pakull \& Mirioni  (\citeyear{Pakull02}) in the source Ho II X-1.

We can therefore conclude that the spectral characteristic of the northern part of the nebula may be better explained by partially radiative shocks. This would be consistent with the evidence described above arguing for a lower pre-shock density in this part of the nebula. We now proceed to investigate this idea in more detail.

\subsection{The Shock Models}
The \HII\ regions identified in Figure \ref{FigvoronoiS26} enable us to determine the appropriate chemical abundance set. We have used the shock/photoionisation code \emph{Mappings  IIIs}, an updated version of the code originally described in Sutherland \& Dopita (\citeyear{Sutherland93}) to generate the models described in L\'opez-S\'anchez et al. (\citeyear{Lopez-Sanchez12}). We then used the line strengths reported in Table \ref{Table1} and the modified strong-line technique also described in the  L\'opez-S\'anchez et al. (\citeyear{Lopez-Sanchez12}) to constrain the abundance in this region of NGC~7793; $Z = 0.5\pm 0.2 Z_{\odot}$. We therefore adopt $Z = 0.5 Z_{\odot}$ for our subsequent analysis of S26.

With this abundance set, we have run a grid of partially-radiative shock models. The shock velocity was varied from $100-220$ km$^{-1}$, and the shock age was constrained by the condition that $\log($\OIII $\lambda5007$/ H$\beta) < 1.0$. We have assumed that the magnetic field pressure and the gas pressure are in equilibrium in the pre-shock gas, with an assumed pre-shock temperature of 10000K. For a pre-shock hydrogen density of 1.0 this corresponds to a transverse component of the magnetic field of $B = 1.68\mu$G.

For all models, we have assumed full pre-ionisation of the gas entering the radiative shock. This assumption will be only valid for shock velocities in excess of 100km~s$^{-1}$, so this is why we have restricted the minimum shock velocity to 100km~s$^{-1}$.

The key parameters and line fluxes of the models is given in Table \ref{Table2}. Column (1) gives, for each value of the shock velocity the corresponding shock age, expressed in terms of $\log(nt)$, where $n$ is the pre-shock hydrogen density, and $t$ is the shock age. All shock models with a given $\log(nt)$ will have very similar emission spectra, modulo collisional de-excitation effects in the density-sensitive lines such as \SII]6717,31\AA. 

Column (2) of  Table \ref{Table2} gives the mean ion-weighted electron density enhancement factor in the zone emitting \SII. This ratio of post-shock to pre-shock electron density increases to a maximum as the gas is compressed by radiative cooling in the post-shock region, before decreasing again as the shocked gas recombines, and the electron density decreases. Note that the computed compression factor is larger for younder shocks. This apparently anomalous result is because the \SII\ emission comes mostly from the peak of the electron density in the partially-radiative shocks, while in the fully radiative shocks the mean \SII\  region electron density is much lower due to the recombination zone in the magnetically supported tail of the shock. The maximum gas density compression factor is higher, and can be approximated by $\rho_1/\rho_0 = 2^{1/2}\cal{M}_{\rm A}$, where $\cal{M}_{\rm A}$ is the Alv\'en Mach Number of the shock.

Column (3) gives the surface brightness of the shock in H$\beta$ for a pre-shock hydrogen density of 1.0~cm$^{-3}$. This will scale in proportion to the pre-shock density.

The remaining columns of Table \ref{Table2} give the strengths of some important emission lines with respect to  H$\beta$. These clearly show that the effect of decreasing shock age is to strengthen \OIII\, weaken \OI\ and \SII\ while the strength of \NII\ is only slightly affected.

\subsection{Comparison with Observations}
\begin{figure*}
    \centering
    \includegraphics[width=0.45\hsize]{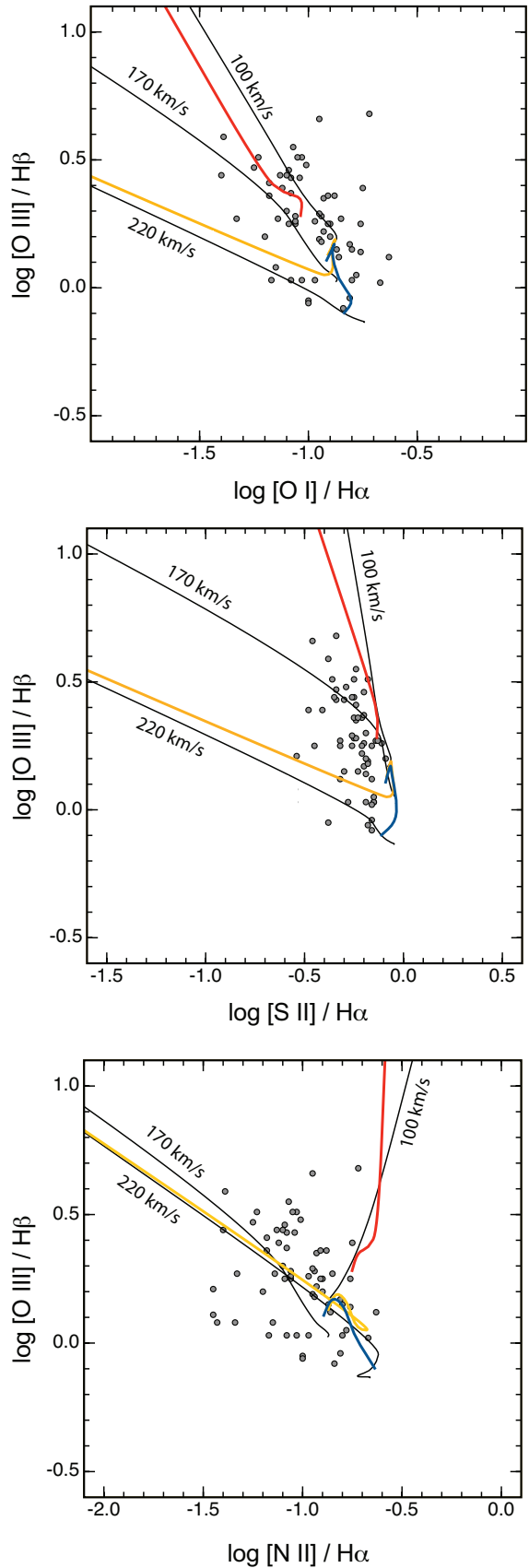}
    \caption{A comparison of the line ratios derived from the partially-radiative shock models described in the text with the observational line ratios in S26. For clarity, a number of models have been omitted. The Red line is an isochrone for $\log(nt) = 11.3$ the yellow line for $\log(nt) = 11.6$ and the blue line for $\log(nt) = 12.6$. These diagnostics are insensitive to age when $\log(nt) > 12.6$.
                 }\label{ratios}
    \end{figure*}

In Figure \ref{ratios} we show the computed line ratios compared with the observed line ratios across the face of S26. The observations imply shock velocities in the computed range of the models, $100-220$ km~s$^{-1}$. The finite age models correctly predict the observed correlation between the (stronger) \OIII/H$\beta$ ratios and the (weaker)  \OI/H$\alpha$ and \SII/H$\alpha$ ratios. As expected there is little correlation between the  \OIII/H$\beta$ and the  \NII/H$\alpha$ ratios. Qualitatively at least, the interpretation of the data as indicating the presence of partially-radiative shocks is confirmed.

  \begin{figure*}
    \centering
    \includegraphics[width=0.6\hsize]{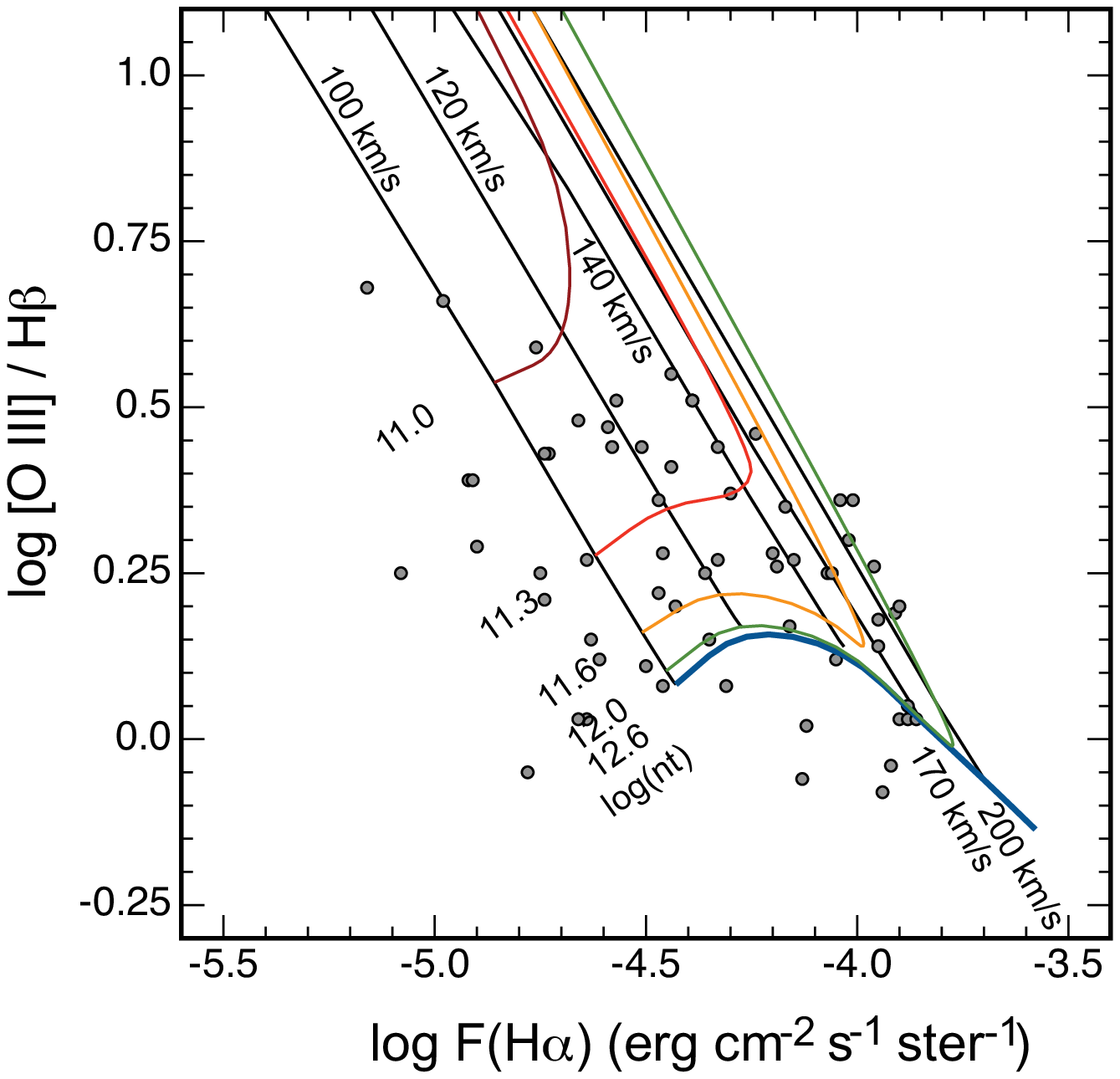}
    \caption{The predicted relationship between the surface brightness in H$\alpha$ and the  \OIII/H$\beta$ ratio as a function of shock velocity and radiative age $\log(nt)$, compared with the observations. Taken at face value, this figure implies shock velocities in the range $80-220$ km$^{-1}$ and radiative ages $\log(nt) > 11$ in S26.
                 } \label{Ha_OIII}
    \end{figure*}
    
The analysis can be made more quantitative by comparing the surface brightness in H$\alpha$ with the observed  \OIII/H$\beta$ ratio. Models predict that for a given shock velocity as the shock becomes less radiative, the  \OIII/H$\beta$ ratio should increase while the H$\alpha$ surface brightness decreases. On the other hand, an increasing shock velocity leads to a higher H$\alpha$ surface brightness as the mechanical luminosity of the shock increases, but the models show little variation in the  \OIII/H$\beta$ ratio with velocity. 

In principle, these facts can be used to independently estimate both the shock velocity, $V_s$ and the radiative age $\log(nt)$. However, we must first remember that in S26, we are looking through both the front and back of the shell, so we see two shocks along the line of sight. With this correction to the predicted flux we obtain the diagnostics shown in Figure \ref{Ha_OIII}. Here, the measured FWHM of the observed H$\alpha$ line has been corrected for the instrumental resolution ($50$km~s$^{-1}$) by subtracting it in quadrature. Figure \ref{Ha_OIII} implies that the lowest shock velocities are of order 80km~s$^{-1}$, while the fastest shocks have $V_s \sim 220$km~s$^{-1}$.

    \begin{figure}
    \centering
    \includegraphics[width=0.8\hsize]{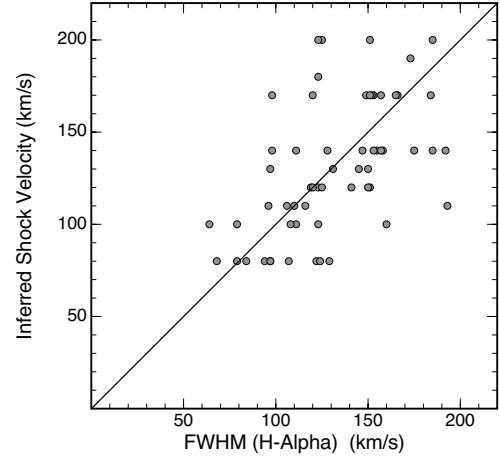}
    \caption{The correlation between the measured FWHM of the H$\alpha$ line and the shock velocity inferred from Figure \ref{Ha_OIII}. The straight line would represent a one to one correlation. The existence of a correlation in this figure helps confirm that the estimated shock velocities are more or less correct.
                 }
    \label{shockcorr}
    \end{figure}

The shock velocities derived by this technique can be compared with the measured FWHM of the  H$\alpha$ line at the corresponding spatial position. We should not expect a one-on-one correlation between these two values, since the FWHM does not necessarily reflect the actual shock velocity. It will be maximised when we observe two shocks moving in opposite directions, and minimised when we observe a shock travelling transversely to the line of sight. Nonetheless, since the radiative shock in general are observed moving into denser clouds, and given the fractal distribution of these we might expect some correlation between FWHM and inferred shock velocity. Indeed there is, as is shown in Figure \ref{shockcorr}. We can conclude that the partially-radiative shock model provides a good description for the velocity width observations of S26.

\subsection{Constraints from He II}
Pakull et al. (\citeyear{pakull10}) and Soria et al. (\citeyear{soria10}) claim (based on narrow-band observations) that there is a strong, extended HeII 4686 nebula, of similar size as the Balmer and [\OIII] nebula, and their estimated He II 4686/H$\beta$ flux ratio is $\sim  0.1$. We do not have data of similar sensitivity in the blue part of the spectrum, but an integration across the nebula gives a flux ratio of He II 4686/H$\beta$ flux $= {0.11}\pm{0.033}$, in good agreement with the imaging data quoted above.

Such a value is consistent with the fast shock velocity,  $V_s \sim 250$km~s$^{-1}$, inferred by Pakull et al. (\citeyear{pakull10}) \emph{cf.}  Dopita \& Sutherland \citeyear{dopita95}, figure 6. The question we need to resolve here is whether such  He II 4686/H$\beta$ flux ratios are also consistent with the partially radiative but slower shocks we infer.

Certainly, strong HeII is produced in the precursor zones of fast shocks (Dopita \& Sutherland \citeyear{dopita95}, \citeyear{dopita96}, Allen et al. \citeyear{allen08}). However, it must be remembered that the extent of the precursors is much greater than the shock itself, so any He II emission will be more spatially extended and more featureless than the shocked shell. 

Surprisingly, the slow shocks are also quite effective in producing He II emission (see Table \ref{Table2}, col. 4). Fully-radiative shocks produce He II 4686/H$\beta$ flux ratios in the range $0.048 - 0.094$ with a peak at about $V_s \sim 120$km~s$^{-1}$. This is consistent with the limit to the He II 4686/H$\beta$ flux ratio that can be placed upon the spectrum shown in Figure 2, taking into account the S/N of this spectrum.

Much higher values can be found in partially-radiative shocks, with values of He II 4686/H$\beta$ $> 0.1$ being readily attainable. Since partially radiative shocks will generally be moving into regions of lower density, and observational test would be to see if the fainter regions of the shell are characterised by higher He II 4686/H$\beta$ flux ratios. Unfortunately the S/N in the blue region of the spectrum is lower than in the red, and a combined spectrum of the faint regions (66-86) is just insufficient to provide a definitive test, since it only constrains He II 4686/H$\beta$ $\lesssim 0.15$.

\subsection{Derived Shock Parameters}
In most regions of S26, the measured \SII\ ratio $\lambda\lambda 6731/6717$ is clearly above the low density limit of 0.665{\footnote{This value follows from the quantum mechanical sum rule for collision strengths between a singlet and a multiplet state; see equation (3.6) of Dopita \& Sutherland (\citeyear{dopita03}), since at low densities the line flux ratio is simply decided by the ratio of the collision strengths from the ground state, corrected by the ratio of energies associated with the individual lines}, even when the measurement errors are taken into account. From Figure \ref{Ha_OIII} we have already estimated the shock velocity, $V_s$ and the radiative age parameter, $\log(nt)$. Thus using the observed  \SII\ $\lambda\lambda 6731/6717$ ratio to infer the \SII\ density, and the appropriate \SII\ compression factor from Table (2) we can infer the pre-shock hydrogen density in each region of the nebula.

The derived shock parameters are listed in Table \ref{Table3}. This gives:
\begin{itemize} 
\item{Col (1): the region number identified in Figure \ref{FigvoronoiS26}.}
\item{Col (2):  the inferred shock velocity from Figure \ref{Ha_OIII}.}
\item{Col (3): the measured FWHM of the H$\alpha$ line.}
\item{Col (4): the age parameter $\log(nt)$ inferred from Figure \ref{Ha_OIII}.}
\item{Col (5): the log of the inferred electron density, $n_e$, from the  \SII\ $\lambda\lambda 6731/6717$ ratio, where values of 1.00 indicate that the ratio is measured at its low density limit.}
\item{Col (6): the inferred pre-shock hydrogen density, $n_0$.}
\item{Col (7): the age of the shock derived from columns (4) and (6), and finally,}
\item{ Col (8): the ram pressure of the shock, $P/k$, derived from columns (2) and (6).}
\end{itemize}

In Figure \ref{P_N_vs_t} we show the correlation between the inferred age of the shock and the density and ram pressure in the shocks. Note that the analysis is insensitive to shock ages $> 10^{13}$s. Broadly speaking, there is an inverse correlation between the pre-shock density and the shock age, and that the pressure in the jet and counter-jet regions is larger than the average in the bubble. The inverse correlation between pre-shock density and shock age is to be expected, since the low density regions can only become radiative over relatively long timescales.

   \begin{figure}
    \centering
    \includegraphics[width=0.8\hsize]{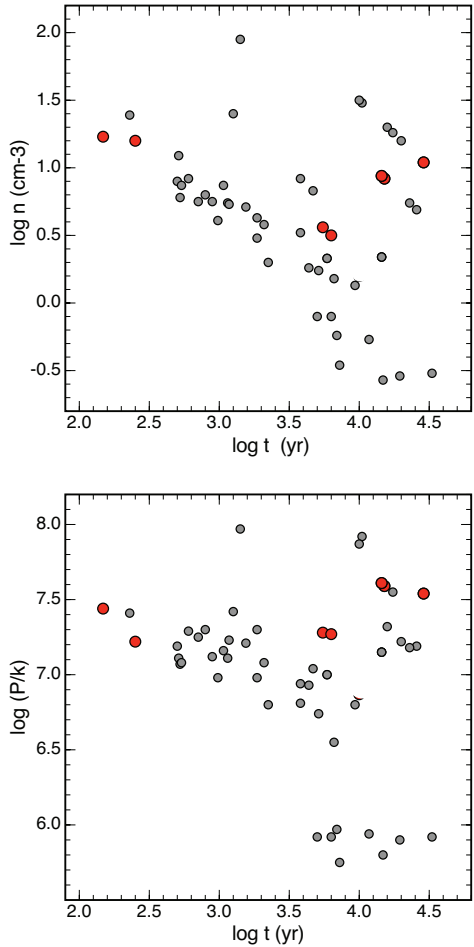}
    \caption{The correlation between inferred age of the shock and either density (upper panel) or the ram pressure (lower panel). The red points on the right represent points in the SE jet and the two red points to the left of these diagrams represent shocks in the NW counter-jet. Note the high pressures in both jets, and the comparative (radiative) youth of the counter-jet shocks.
                 }
    \label{P_N_vs_t}
    \end{figure}

\section{Global Parameters of S26}
The theory of relativistic jet-driven bubbles was developed in the context of the double lobe radio sources in active galactic nuclei (AGN) over many years by Scheuer (\citeyear{Scheuer74}), Blandford \& Rees (\citeyear{Blandford74}), Rawlings \& Saunders (\citeyear{Rawlings91}), Kaiser \& Alexander (\citeyear{Kaiser99}), Begelman  (\citeyear{Begelman96}) and Bicknell et al. (\citeyear{Bicknell97}). Zealey, Dopita \& Malin (\citeyear{Zealey80}) developed the theory to apply to the case of the micro-quasar SS433. In summary, the relativistic jet is shocked when it starts to burrow into the ISM and the overpressure of the backflow carves out a cavity or cocoon around the jet. The back-flowing gas then inflates a larger bubble between the two jets, which behaves like a mass-loss bubble. The detailed physics is given in Dopita \& Sutherland (\citeyear{dopita03}). Define the $r-$axis to be in the direction along the jets, and the $z-$axis to be the perpendicular direction. If the mean pressure in the cocoon is $P$, then towards the head of the cocoon (in the $r$ direction) the mean pressure is higher, because this is defined by the small area,$ A$, over which the termination
shock is jittering. Here we can take the pressure as a factor $\zeta$ times the average lobe pressure where, typically  $\zeta \sim$ 2 - 10. The velocity of advance of the jet, and the velocity of the wall shocks surrounding the cavity excavated by the jet are then given by:
\begin{equation}
\frac{dr}{dt} =\left[\frac{\beta\dot {E_j}}{\rho cA}\right]^{1/2} \sim \zeta^{1/2}\left[\frac{P}{\rho}\right]^{1/2} \label{eqn1}
\end{equation}
and 
\begin{equation}
\frac{dz}{dt} =\left[\frac{P}{\rho}\right]^{1/2}  \label{eqn2}
\end{equation}
respectively where ${\rho}$ is the density of the pre-shock ISM surrounding the jet head, and $\dot {E_j}$ is the mechanical energy delivered to a single jet moving at  a velocity $\beta c$, and having an area $A$.

For the larger bubble between the jets, Zealey, Dopita \& Malin (\citeyear{Zealey80})  showed that the mass-loss bubble formula for the bubble radius along the minor axis, $z$,  given by Castor et al (\citeyear{Castor75}) or Weaver et al. (\citeyear{Weaver77}) applies:
\begin{equation}z=\alpha ^{1/5}\left( \frac{\stackrel{.}{E}_{j}}{\rho_{z}}\right) ^{1/5}t^{3/5}   \label{eqn3}
\end{equation}
where $\stackrel{.}{E}_{j}$ is now understood to be the combined energy flux of both jets, $\rho_z$ is the density along the minor axis, and $\alpha$ is a constant. For a bubble containing thermal plasma with a filling factor $\phi$,  $\alpha = \left( {125}/{154\pi \phi}\right)$ so that $\alpha^{1/5}=0.76\phi^{-1/5}$, while for a bubble entirely filled by relativistic plasma with $\gamma = 4/3$, $\alpha = ({125}/{224\pi \phi })$ so that $\alpha^{1/5} = 0.708\phi^{-1/5}$. The latter case (with $\phi \sim 1.0$) is more likely to be valid in the case of S26, since the extent of the non-thermal radio emission is quite similar to that of the optical shell.\footnote{We are grateful to the referee for reminding us of the subtle distinction between the thermally and the relativistically filled cases.} The age of the bubble, $\tau$, can then be directly inferred from the radius of the bubble, $z$, and the expansion velocity in the $z-$direction;  $\tau = (3/5)(z/v_z)$.

The density, pressure and shock velocity in the cocoon around the SE jet can be obtained by averaging the values estimated for Spectrum numbers 24, 25, 28, 29, 32, 33, and 34. This gives $v_s(z) = 165$km s$^{-1}$, $P=6.0\times10^{-9}$dynes cm$^{-2}$ and $\rho = 2.2\times10^{-23}$g cm$^{-3}$. The likely value of  $\zeta$ can be estimated by comparing the rate of advance of the jets given by equation \ref{eqn2} with the observations. With $\zeta \sim 4$ the rate of advance of the SE jet of 330 km s$^{-1}$ is reproduced. Applying a similar analysis to the NW counter-jet (spectrum numbers 79 and 86) gives $v_s(z) = 130$km s$^{-1}$, $P=2.5\times10^{-9}$dynes cm$^{-2}$ , $\rho = 1.5\times10^{-23}$g cm$^{-3}$ and a current rate of advance of the NW jet $v_s(r)$, of 260 km s$^{-1}$, which is also in good agreement to the observations. We may conclude that  $\zeta \sim 4$. Finally, given that we would expect the the pressure at the termination shocks of the jets to fall off as the inverse square of the distance, the ratio of the measured pressures in the two jets should be in proportion to the inverse square of the distance. For S26 this implies a pressure ratio of $\sim2.2$ which compares well to the observed ratio; 2.4.

The velocity of expansion of the overall mass-loss bubble can be also estimated using the mean of spectra 66, 67, 70 and 72. We find $v_s(z) = 132$km s$^{-1}$, $P=2.0\times10^{-9}$dynes cm$^{-2}$ , $\rho_z = 1.15\times10^{-23}$g cm$^{-3}$. The ratio of $v_s(r)/v_s(z) = 2.5$ compares well with the ratio of the physical size in the $r$ and $z$ directions, $R/Z = $175pc~/~92pc = 1.9, showing that the expansion velocity estimates are reasonably consistent with each other. 

The age of S26 can now be estimated as $\tau = (3/5)(z/v_z) = 1.3 \pm 0.15 \times 10^{13}$s or $4.1 \pm0.5 \times10^5$yr. This is somewhat older than the $\sim2\times10^5$yr estimated by \cite{pakull10} thanks to the lower shock velocity estimated here. 

The jet energy flux can now be obtained from equation \ref{eqn3}. We find $\stackrel{.}{E}_{j} = 5.6^{+2.0}_{-1.5}\times10^{40}$ erg s$^{-1}$, identical to the value estimated by \cite{pakull10} from the H$\beta$ luminosity. However, they used the theory of fast shocks from \cite{allen08}, whereas slow shock theory is more appropriate for the vast majority of the shocks in S26. For these, the conversion factor between H$\beta$  (or H$\alpha$) flux and total luminosity is lower, and is given by \cite{Rich10}. For shock velocities above 120 and less than 200km s$^{-1}$, the conversion factor is virtually constant at about 80 (H$\alpha$ to total luminosity). We therefore estimate the corrected total luminosity of the radiative shocks in S26 to be $L_{rad}= 2.6\times10^{40}$ erg s$^{-1}$.  Using the theory appropriate for a fully filled relativistic bubble in the fully radiative phase, $\stackrel{.}{E}_{j} =(112/25)L_{rad} =1.0\times10^{41}$ erg s$^{-1}$ ($7.0\times10^{40}$ erg s$^{-1}$ if filled with thermal plasma). Combining these two approaches, we conclude that the jet energy flux lies in the range $(4-10)\times10^{40}$ erg s$^{-1}$, in excellent agreement with that estimated by \cite{pakull10}.

The jet energy flux corresponds to an Eddington Luminosity for a $200-700$M$_{\odot}$ Black Hole. However, it is comparable with the X-ray luminosities of the most hyper-luminous stellar sources (Swartz et al., \citeyear{Swartz11}), strongly suggesting that S26 belongs to the same population of stellar Black Holes as the ULX sources. The difference is presumably that S26 is currently attempting to swallow a stellar companion through Roche Lobe accretion so that the X-ray luminosity has been transformed to jet energy flux.

By substituting the derived jet energy flux into equation \ref{eqn1} we can estimate the area $A$ of the shocks. We do not have a good estimate of the relativistic $\beta$, so retain this factor explicitly to  estimate $A \sim 0.38\beta^{-1}$pc$^2$. If we assume $\beta$ to be similar to SS433 ($\beta \sim 1/3$), then the jet covers about 1.1pc$^2$, implying a collimation angle of about 0.25 degrees. 

If the X-ray hotspots are indeed marginally resolved, as claimed by \cite{pakull10}, then they cover $\sim 20$pc and subtend an angle of 6 degrees at the central source. Therefore, either the precession angle or ``jittering'' of the jets is within a cone opening angle of 6 degrees, or else the jet is de-collimated by the same amount. This latter condition would require a very low jet speed, $\beta \sim 0.02$, of a jet velocity of $\sim7000$kms$^{-1}$. Whilst not excluded this seems unlikely, as it would require that a great deal of matter has been processed by the jet. This mass is given by $M=\tau \stackrel{.}{E}_{j}/(\beta c)^2$. For $\beta =1/3$ we obtain $M \sim 4$M$_{\odot}$, while for $\beta = 0.02$, the mass needed to be processed rises to over 1000M$_{\odot}$. Clearly the jet energy flux cannot be maintained over the lifetime of the bubble unless the jets are fully relativistic. This also suggests that here we are dealing with a bubble filled with relativistic rather than thermal gas. The total energy injected to the bubble over the lifetime of the bubble is given by $E_{tot} =  \stackrel{.}{E}_{j} \tau \sim 8\times10^{53}$ ergs.

\section{Conclusions}
We have shown that the radiative properties of the shell of NGC 7793 - S26 can be understood in the context of a model in which the shell is still becoming fully-radiative in its NW portions, due to a large-scale density gradient in the ISM. Although the age of the oldest shocks cannot be reliably measured we can set an age limit comparable with the inferred age of the shell, $4.1\times10^5$yr, and we can infer a jet advance velocity in the SE jet of 330 km~s$^{-1}$.

We concur with \cite{pakull10} that S26 represents the ``missing link'' between the ultra-luminous X-ray sources and their energetic bubbles, and the less luminous Galactic source SS433 and its nebula, W50. Indeed, the lower jet luminosity we infer here,  $\stackrel{.}{E}_{j} = (4-10)\times10^{40}$ erg s$^{-1}$ is comparable with the X-ray luminosities of the most hyper-luminous stellar sources \cite{Swartz11}. Unlike SS433, the jets in S26 undergo very little precession; over not more than $\pm3$ degrees.

In conclusion, the central source of S26 is almost certainly a Black Hole of the typical mass which characterizes the ULX population in a close binary with a more normal star. Given that S26 has processed at least 4M$_{\odot}$ through the jets already, the companion is likely to be an intermediate-mass star which is in the process of being consumed by its unfriendly companion.

\begin{table*}
\caption{Selected line fluxes (at rest wavelengths) and log ratios for N7793-S26 and the nearby \HII\ region to its south as shown in Fig. 3.  Note in the case of  \SII / H$\alpha$, a ratio greater than 0.4 (the classical definition for a region dominated by shock emission; Lasker 1977; Mathewson et al. 1983, 1984, 1985) corresponds to a log value greater than --0.4.
     }\label{Table1}
\centering
\resizebox{!}{8.3cm} {\begin{tabular}{cccccccccccccccc}
%\multicolumn{19}{c}{{\bf Position Angle 0\D}}\\
\noalign{\smallskip} \hline\hline\noalign{\smallskip}
% (1)& (2)  & (3) & (4)    & (5) &  (6)      &  (7)   & (8)& (9)&(10) \\
&    \multicolumn{10}{c}{Emission Line Flux 
($\times 10^{-15}$ ergs cm$^{-2}$ s$^{-1}$)} &\multicolumn{5}{c}{Log}\\
 Spectrum  
  & 4861 \AA &4959 \AA &
5007 \AA& 6300 \AA &6364 \AA&6550 \AA& 6563 \AA& 6585 \AA& 6716 \AA& 6731 \AA& 5007 \AA / 4861 \AA  &
6300 \AA /  6563 \AA&
6585 \AA /  6563 \AA& (6716 \AA + 6731 \AA) / 6563 \AA &H$\alpha$ Surface Brightness \\
 Number       & H$\beta$&\OIII&\OIII&\OI&\OI&\NII&H$\alpha$&\NII&\SII&\SII & \OIII / H$\beta$ & \OI / H$\alpha$ & \NII / H$\alpha$  &\SII / H$\alpha$&(ergs s$^{-1}$ cm$^{-2}$ sr$^{-1}$)\\
\noalign{\smallskip} \hline \noalign{\smallskip}
5&2.3&0.8&2.8&0.3&&1.8&7.5&0.9&1.0&0.7&0.08 &--1.34&--0.91&--0.63&--4.31\\
 6&2.4&1.0&2.9&0.3&&2.2&6.9&1.2&1.1&0.9&0.08 &--1.43&--0.77&--0.55&--4.46\\
 10&1.9&1.0&2.4&0.3&&2.6&7.1&1.1&1.0&0.7&0.11&--1.45 &--0.80&--0.60&--4.50\\
 12&1.7&0.8&1.8&0.4&&2.9&6.5&0.9&1.0&0.8&0.03 &--1.17&--0.84&--0.55&--4.64\\
 13&1.7&1.0&2.8&0.2&&4.3&7.1&0.8&1.2&0.8&0.21&--1.45&--0.95&--0.54&--4.74\\
  \noalign{\smallskip} \hline \noalign{\smallskip} \\
   15&1.1&0.6&1.3&0.2&&1.9&3.2&&0.9&0.7&0.08&--1.15 & &--0.30&--4.79\\
  19&1.0&0.6&0.9&0.2&&1.3&2.2&0.2&0.5&0.4&--0.05&--1.00 &--0.97&--0.38&--4.78\\
  20&0.9&0.3&0.8&0.3&&0.4&2.6&0.4&0.9&0.9&--0.06&--1.00&--0.81&--0.18&--4.13\\
  21&1.0&0.4&1.0&0.2&0.2&1.2&2.9&0.2&0.7&0.8&0.03 &--1.08&--1.17&--0.28&--4.66\\
 23&0.6&0.4&0.8&0.2&&0.7&1.7&0.3&0.5&0.3&0.12&--0.86 &--0.75&--0.32&--4.61\\
 24&0.3&0.1&0.3&0.2&0.1&0.2&1.4&0.2&0.5&0.4&--0.04&--0.81&--0.85&--0.16&--3.92\\
  25&0.9&0.3&0.7&0.4&0.2&0.3&2.7&0.4&1.0&0.9&--0.08&--0.84 &--0.84&--0.16&--3.94\\
   26&0.8&0.5&1.1&0.3&0.2&0.7&1.7&0.3&0.5&0.4&0.15&--0.80 &--0.78&--0.30&--4.63\\
   27&0.5&0.3&0.7&0.2&0.1&0.4&1.6&0.4&0.5&0.4&0.15&--0.87&--0.65&--0.23&--4.35\\
 28&0.4&0.1&0.4&0.2&0.1&0.2&1.5&0.2&0.6&0.5&0.03 &--0.97&--0.84&--0.15&--3.90\\
29&0.4&0.2&0.5&0.2&0.1&0.1&1.5&0.2&0.6&0.5&0.03&--0.80 &--0.80&--0.15&--3.88\\
30&0.2&0.1&0.3&0.2&0.0&0.1&0.9&0.1&0.3&0.3&0.02 &--0.67&--0.93&--0.16&--4.12\\
31&0.6&0.4&1.0&0.3&0.2&0.7&2.4&0.4&0.9&0.6&0.22 &--0.93&--0.78&--0.21&--4.47\\
 32&0.4&0.1&0.5&0.2&0.1&0.2&1.6&0.2&0.6&0.5&0.03&--1.03 &--0.90&--0.19&--3.86\\
 33&0.5&0.2&0.5&0.3&0.1&0.1&1.6&0.3&0.6&0.5&0.05&--0.78 &--0.77&--0.15&--3.88\\
 34&0.3&0.2&0.4&0.2&0.1&0.2&1.1&0.2&0.4&0.3&0.12 &--0.63&--0.84&--0.16&--4.05\\
35&0.4&0.3&0.6&0.3&0.1&0.2&1.6&0.3&0.6&0.4&0.17 &--0.81&--0.74&--0.20&--4.16\\
36&0.7&0.4&1.1&0.3&0.2&0.3&2.9&0.5&1.1&0.8&0.19&--0.95 &--0.77&--0.18&--3.91\\
37&0.4&0.2&0.5&0.2&0.1&0.2&1.3&0.2&0.5&0.4&0.14 &--0.76&--0.81&--0.19&--3.95\\
38&0.5&0.4&0.9&0.2&0.1&0.3&1.1&0.2&0.3&0.2&0.27&--0.85 &--0.81&--0.32&--4.33\\
42&0.5&0.3&1.0&0.3&0.2&0.5&2.1&0.3&0.7&0.5&0.25&--0.90 &--0.83&--0.26&--4.36\\
43&0.4&0.2&0.6&0.1&0.1&0.2&1.3&0.3&0.5&0.4&0.18&--0.94&--0.71&--0.18&--3.95\\
44&0.8&0.5&1.3&0.4&0.2&0.3&3.0&0.5&1.1&0.8&0.20&--0.90 &--0.77&--0.19&--3.90\\
45&0.5&0.4&1.0&0.2&0.1&0.3&1.5&0.3&0.5&0.4&0.28&--0.94 &--0.66&--0.25&--4.20\\
46&1.1&0.8&2.4&0.3&0.4&1.5&4.4&0.8&1.5&1.2&0.36 &--1.18&--0.73&--0.21&--4.47\\
47&0.3&0.2&0.5&0.1&0.1&0.1&1.0&0.2&0.4&0.3&0.25&--1.06 &--0.67&--0.16&--4.07\\
48&0.3&0.2&0.6&0.1&&0.1&1.1&0.2&0.4&0.3&0.30 &--1.10&--0.84&--0.19&--4.02\\
49&0.4&0.3&0.7&0.1&0.1&0.2&1.3&0.2&0.5&0.3&0.26 &--0.97&--0.73&--0.20&--3.96\\
50&0.6&0.5&1.4&0.2&0.1&0.3&1.6&0.3&0.5&0.4&0.35 &--0.93&--0.72&--0.24&--4.17\\
51&0.5&0.4&0.9&0.1&0.1&0.2&1.5&0.3&0.7&0.5&0.26&--1.06 &--0.66&--0.11&--4.19\\
52&0.6&0.3&1.1&0.1&&0.3&1.6&0.2&0.6&0.5&0.27&--1.14 &--0.86&--0.14&--4.15\\
53&0.7&0.5&1.3&0.3&0.1&0.3&2.0&0.4&0.7&0.5&0.25 &--0.91&--0.74&--0.21&--4.06\\
54&0.4&0.3&0.8&0.1&0.1&0.2&1.1&0.2&0.4&0.3&0.36&--0.88 &--0.83&--0.25&--4.01\\
55&0.3&0.2&0.8&0.1&0.0&0.2&1.1&0.2&0.4&0.3&0.36&--0.91 &--0.81&--0.23&--4.04\\
56&0.8&0.6&1.5&0.1&0.1&0.5&1.6&0.3&0.5&0.4&0.28&--1.06 &--0.74&--0.24&--4.46\\
60&1.4&0.8&2.2&0.2&&0.9&3.5&0.6&1.6&1.2&0.20&--1.20 &--0.80&--0.09&--4.43\\
61&0.6&0.6&1.3&0.1&&0.4&1.8&0.3&0.6&0.5&0.37 &--1.08&--0.74&--0.20&--4.30\\
62&0.4&0.4&1.2&0.1&&0.3&1.3&0.2&0.5&0.3&0.46&--1.09 &--0.78&--0.20&--4.24\\
63&1.4&1.2&2.6&0.1&&1.2&2.9&0.4&1.3&0.9&0.27 &--1.33&--0.89&--0.13&--4.64\\
64&0.9&0.9&2.4&0.1&&0.6&2.7&0.3&0.9&0.7&0.44&--1.40 &--0.92&--0.25&--4.33\\
65&0.6&0.7&1.9&0.1&&0.8&2.1&0.4&0.6&0.4&0.47 &--1.25&--0.72&--0.34&--4.59\\
66&0.7&1.0&2.6&0.2&&0.8&2.5&0.4&0.8&0.6&0.55 &--1.07&--0.81&--0.24&--4.44\\
67&0.8&0.6&1.4&0.2&&1.1&2.1&0.5&0.6&0.5&0.25&--1.09 &--0.65&--0.32&--4.75\\
69&1.3&1.1&2.6&0.4&&2.4&3.2&0.4&0.9&0.9&0.29&--0.95 &--0.88&--0.26&--4.90\\
70&0.9&1.0&2.9&0.3&&0.8&2.9&0.4&0.7&0.6&0.51 &--1.03&--0.87&--0.36&--4.39\\
71&1.0&1.1&3.1&0.2&&0.9&2.5&0.4&0.9&0.7&0.51 &--1.05&--0.80&--0.18&--4.57\\
72&0.6&0.7&1.9&0.1&0.1&0.5&1.9&0.4&0.6&0.5&0.51&--1.23 &--0.71&--0.25&--4.39\\
73&0.7&0.8&2.1&0.3&&1.4&2.8&0.7&0.8&0.7&0.48 &--1.01&--0.62&--0.29&--4.66\\
74&0.5&1.0&2.3&0.8&&5.7&4.0&0.7&1.1&0.7&0.68&--0.72 &--0.76&--0.34&--5.16\\
75&0.9&1.7&3.5&0.1&&2.0&3.5&1.0&0.8&0.6&0.59&--1.39&--0.56&--0.38&--4.76\\
76&0.9&0.8&2.3&0.2&0.1&0.7&2.6&0.4&0.9&0.6&0.41&--1.18 &--0.76&--0.24&--4.44\\
77&0.7&0.7&1.9&0.2&&1.5&2.6&0.5&0.8&0.6&0.43&--1.08&--0.72&--0.30&--4.73\\
78&0.7&1.2&3.1&0.3&&2.7&2.9&0.3&0.5&0.5&0.66&--0.95 &--0.95&--0.46&--4.98\\
79&1.6&1.6&4.3&0.3&&1.4&4.4&0.7&1.3&1.1&0.44 &--1.10&--0.79&--0.26&--4.51\\
80&1.2&1.4&3.3&0.3&&1.9&3.7&0.4&1.0&0.7&0.43 &--1.04&--0.94&--0.34&--4.74\\
82&2.3&1.4&4.1&1.0&&7.0&5.7&0.9&1.2&0.8&0.25 &--0.76&--0.79&--0.45&--5.08\\
83&1.8&1.7&4.4&0.8&&3.6&4.2&0.6&0.8&0.6&0.39 &--0.75&--0.84&--0.48&--4.92\\
84&1.6&1.3&4.3&0.3&&1.7&4.6&0.7&1.1&0.9&0.44 &--1.13&--0.80&--0.35&--4.58\\
86&2.6&2.2&6.4&0.4&&4.1&5.3&1.1&1.2&0.8&0.39 &--1.12&--0.67&--0.41&--4.91\\
 \noalign{\smallskip} \hline \noalign{\smallskip} \\
\end{tabular}}
\end{table*}

\begin{table*}
\caption{The finite age (partially radiative) shock models. All have a metallicity $0.5 Z_{\odot}$. Line fluxes are given with respect to H$\beta = 1.0$.
     }\label{Table2}
\centering
\resizebox{!}{8.3cm} {\begin{tabular}{ccccccccc}
\noalign{\smallskip} \hline\hline\noalign{\smallskip}
(1) & (2) & (3) & (4) & (5) & (6) & (7) & (8) & (9)\\
$\log (nt)$ & \SII\ Comp.& $\log F H_{\beta}$ & He II & \OIII\ & \OI\ & H$\alpha$ & \NII\ & \SII\ \\
cm$^{3}$s & Factor & erg cm$^{2}$s$^{-1}$sr$^{-1}$ & 4686 & 5007 & 6300 & 6563 & 6584 & 6717+31 \\
\noalign{\smallskip} \hline \noalign{\smallskip}
$V_s = 100$ kms$^{-1}$ &  &  &  &  &  &  &   \\
 12.6021 & 12.0 & -5.2206 & 0.0481 &1.2090 & 0.3544 & 3.0970 & 0.3728 & 2.3861 \\
 12.3010 & 12.8 & -5.2251 & 0.0486 &1.2210 & 0.3581 & 3.0860 & 0.3767 & 2.4111 \\
 12.0000 & 14.7 & -5.2410 & 0.0504 & 1.2670 & 0.3707 & 3.0770 & 0.3908 & 2.4910 \\
 11.7782 & 16.9 & -5.2640 & 0.0531 & 1.3360 & 0.3767 & 3.0760 & 0.4119 & 2.5240 \\
 11.6021 & 19.3 & -5.2991 & 0.0574 & 1.4480 & 0.3606 & 3.0820 & 0.4452 & 2.4720 \\
 11.3010 & 23.8 & -5.4150 & 0.0746 &1.8910 & 0.2869 & 3.1210 & 0.5510 & 2.2865 \\
 11.0000 & 28.2 & -5.6755 & 0.1335 & 3.4450 & 0.2082 & 3.2780 & 0.7975 & 2.2044 \\
\noalign{\smallskip} \hline \noalign{\smallskip}
$V_s = 120$ kms$^{-1}$ &  &  &  &  &  &  &   \\
 12.6021 & 15.8 & -5.0593 & 0.0937 & 1.4750 & 0.3956 & 3.0670 & 0.4123 & 2.5420 \\
 12.3010 & 17.0 & -5.0625 & 0.0944 & 1.4810 & 0.3983 & 3.0590 & 0.4148 & 2.5590 \\
 12.0000 & 18.8 & -5.0748 & 0.0971 & 1.5240 & 0.4092 & 3.0500 & 0.4268 & 2.6250 \\
 11.7782 & 21.9 & -5.0938 & 0.1013 & 1.5920 & 0.4133 & 3.0480 & 0.4457 & 2.6430 \\
 11.6021 & 24.3 & -5.1219 & 0.1079 & 1.6980 & 0.3934 & 3.0520 & 0.4741 & 2.5690 \\
 11.3010 & 30.6 & -5.2461 & 0.1428 & 2.2610 & 0.2915 & 3.0870 & 0.5829 & 2.2906 \\
 11.0000 & 35.5 & -5.4814 & 0.2430 & 3.8860 & 0.2101 & 3.2050 & 0.7704 & 2.1584 \\
\noalign{\smallskip} \hline \noalign{\smallskip}
$V_s = 140$ kms$^{-1}$ &  &  &  &  &  &  &   \\
 12.6021 & 22.8 & -4.8116 & 0.0656 & 1.3780 & 0.3733 & 3.0160 & 0.4738 & 2.5500 \\
 12.3010 & 23.6 & -4.8132 & 0.0659 & 1.3830 & 0.3747 & 3.0110 & 0.4755 & 2.5600 \\
 12.0000 & 26.2 & -4.8215 & 0.0671 & 1.4100 & 0.3815 & 3.0030 & 0.4847 & 2.6050 \\
 11.7782 & 29.8 & -4.8375 & 0.0695 & 1.4630 & 0.3810 & 3.0010 & 0.5026 & 2.6060 \\
 11.6021 & 34.1 & -4.8710 & 0.0749 & 1.5800 & 0.3463 & 3.0040 & 0.5391 & 2.4950 \\
 11.3010 & 41.5 & -5.0405 & 0.1099 & 2.3340 & 0.2346 & 3.0390 & 0.6505 & 2.2247 \\
 11.0000 & 46.8 & -5.5030 & 0.3107 & 6.7670 & 0.1524 & 3.2840 & 0.9212 & 2.0544 \\
\noalign{\smallskip} \hline \noalign{\smallskip}
$V_s = 170$ kms$^{-1}$ &  &  &  &  &  &  &   \\
12.6021 & 30.3 & -4.6190 & 0.0553 & 1.0550 & 0.3997 & 2.9950 & 0.5285 & 2.7290 \\
12.3010 & 31.4 & -4.6193 & 0.0554 & 1.0590 & 0.4010 & 2.9900 & 0.5302 & 2.7380 \\
12.0000 & 35.1 & -4.6286 & 0.0564 & 1.0770 & 0.4066 & 2.9850 & 0.5394 & 2.7720 \\
 11.7782 & 40.0 & -4.6414 & 0.0581 & 1.1140 & 0.3935 & 2.9830 & 0.5575 & 2.7180 \\
 11.6021 & 46.7 & -4.6853 & 0.0637 & 1.2330 & 0.3312 & 2.9870 & 0.6044 & 2.5350 \\
 11.3010 & 56.8 & -5.0327 & 0.1380 & 2.7410 & 0.1890 & 3.0660 & 0.7340 & 2.1844 \\
\noalign{\smallskip} \hline \noalign{\smallskip}
$V_s = 200$ kms$^{-1}$ &  &  &  &  &  &  &   \\
 12.6021 & 34.5 & -4.4695 & 0.0589 & 0.8596 & 0.5061 & 2.9940 & 0.6038 & 2.8170 \\
 12.3010 & 37.2 & -4.4731 & 0.0594 & 0.8650 & 0.5102 & 2.9880 & 0.6088 & 2.8410 \\
 12.0000 & 45.7 & -4.4870 & 0.0608 & 0.8948 & 0.4868 & 2.9860 & 0.6281 & 2.7320 \\
 11.7782 & 62.1 & -4.5893 & 0.0728 & 1.1320 & 0.3173 & 2.9970 & 0.7081 & 2.2227 \\
\noalign{\smallskip} \hline \noalign{\smallskip}
$V_s = 220$ kms$^{-1}$ &  &  &  &  &  &  &   \\
 12.6021 & 37.8 & -4.3539 & 0.0570 & 0.7310 & 0.5367 & 2.9870 & 0.6477 & 2.6650 \\
 12.4771 & 38.9 & -4.3551 & 0.0571 & 0.7332 & 0.5381 & 2.9840 & 0.6494 & 2.6730 \\
 12.3010 & 42.6 & -4.3558 & 0.0579 & 0.7395 & 0.5417 & 2.9820 & 0.6550 & 2.6860 \\
 12.1761 & 47.0 & -4.3640 & 0.0592 & 0.7384 & 0.5293 & 2.9810 & 0.5293 & 2.6210 \\
 12.0000 & 59.1 & -4.3875 & 0.0717 & 0.7900 & 0.4315 & 2.9820 & 0.6929 & 2.2801 \\
 11.9031 & 72.9 & -4.5020 & 0.1228 & 1.0280 & 0.2858 & 2.9910 & 0.7448 & 1.8693 \\
 \noalign{\smallskip} \hline \noalign{\smallskip} \\
 
\end{tabular}}
\end{table*}

\begin{table*}
\caption{The inferred shock parameters for each region of S26 defined by Figure \ref{FigvoronoiS26}.
     }\label{Table3}
\centering
\resizebox{!}{8.3cm} {\begin{tabular}{cccccccc}
\noalign{\smallskip} \hline\hline\noalign{\smallskip}
(1) & (2) & (3) & (4) & (5) & (6) & (7) & (8) \\
Spectrum & $V_s $& FWHM & $\log(nt)$ & $\log n_e$ & $\log n_0$ & $\log t $ & $\log(P/k)$ \\
Number & (km s$^{-1}$) & (km s$^{-1}$) & (cm$^{-3}$s) & (cm$^{-3}$) & (cm$^{-3}$) & (yrs) & (cm$^{-3}$K) \\
 \noalign{\smallskip} \hline \noalign{\smallskip} \\
 15 & 80 & 122 & $>13.0$ & 2.20 & 1.20 & 4.30 & 7.22 \\
19 & 80 & 97 & $>13.0$ & 2.30 & 1.30 & 4.20 & 7.32 \\
20 & 130 & 150 & $>13.0$ & 2.76 & 1.48 & 4.02 & 7.92 \\
21 & 80 & 129 & 12.6 & 2.95 & 1.95 & 3.15 & 7.97 \\
23 & 80 & 94 & 12.0 & 2.00 & 0.92 & 3.58 & 6.94 \\
24 & 140 & 147 & $>13.0$ & 2.40 & 1.04 & 4.46 & 7.54 \\
25 & 140 & 158 & $>13.0$ & 2.60 & 1.04 & 4.46 & 7.54 \\
26 & 80 & 97 & 12.0 & 2.40 & 1.40 & 3.10 & 7.42 \\
27 & 110 & 106 & $>13.0$ & 2.40 & 1.26 & 4.24 & 7.55 \\
28 & 170 & 153 & 12.6 & 2.40 & 0.92 & 4.18 & 7.59 \\
29 & 170 & 152 & 12.6 & 2.40 & 0.92 & 4.18 & 7.59 \\
30 & 120 & 119 & $>13.0$ & 2.70 & 1.50 & 4.00 & 7.87 \\
31 & 100 & 111 & 12.0 & 2.00 & 0.83 & 3.67 & 7.04 \\
32 & 170 & 166 & 12.6 & 2.40 & 0.94 & 4.16 & 7.61 \\
33 & 170 & 149 & 12.6 & 2.40 & 0.94 & 4.16 & 7.61 \\
34 & 140 & 111 & 12.6 & 2.05 & 0.69 & 4.41 & 7.19 \\
35 & 130 & 131 & 12.6 & 2.00 & 0.74 & 4.36 & 7.18 \\
36 & 170 & 165 & 11.7 & 1.80 & 0.20 & 4.00 & 6.87 \\
37 & 180 & 123 & 11.8 & 2.20 & 0.56 & 3.74 & 7.28 \\
38 & 130 & 97 & 11.5 & 1.00 & -0.52 & 4.52 & 5.92 \\
42 & 120 & 123 & 11.5 & 1.60 & 0.18 & 3.82 & 6.55 \\
43 & 190 & 173 & 11.8 & 2.20 & 0.50 & 3.80 & 7.27 \\
44 & 200 & 151 & 11.8 & 1.80 & 0.01 & 4.29 & 6.82 \\
45 & 140 & 98 & 11.4 & 2.20 & 0.58 & 3.32 & 7.08 \\
46 & 120 & 141 & 11.3 & 2.23 & 0.74 & 3.06 & 7.11 \\
47 & 170 & 184 & 11.6 & 2.00 & 0.33 & 3.77 & 7.00 \\
48 & 200 & 185 & 12.0 & 2.00 & 0.34 & 4.16 & 7.15 \\
49 & 170 & 151 & 11.6 & 2.00 & 0.33 & 3.77 & 7.00 \\
50 & 170 & 98 & 11.4 & 2.35 & 0.63 & 3.27 & 7.30 \\
51 & 140 & 192 & 11.5 & 1.80 & 0.24 & 3.71 & 6.74 \\
52 & 140 & 185 & 11.4 & 2.30 & 0.71 & 3.19 & 7.21 \\
53 & 170 & 157 & 11.6 & 1.80 & 0.13 & 3.97 & 6.80 \\
54 & 200 & 125 & 12.0 & 2.00 & 0.34 & 4.16 & 7.15 \\
55 & 200 & 123 & 12.0 & 2.00 & 0.34 & 4.16 & 7.15 \\
56 & 110 & 96 & 11.4 & 2.35 & 0.87 & 3.03 & 7.16 \\
60 & 110 & 193 & 11.6 & 2.00 & 0.52 & 3.58 & 6.81 \\
61 & 140 & 175 & 11.3 & 2.35 & 0.73 & 3.07 & 7.23 \\
62 & 170 & 120 & 11.4 & 2.00 & 0.26 & 3.64 & 6.93 \\
63 & 100 & 160 & 11.3 & 1.10 & -0.27 & 4.07 & 5.94 \\
64 & 140 & 155 & 11.3 & 2.10 & 0.48 & 3.27 & 6.98 \\
65 & 120 & 120 & 11.1 & 1.00 & -0.57 & 4.17 & 5.80 \\
66 & 140 & 157 & 11.2 & 1.95 & 0.30 & 3.35 & 6.80 \\
67 & 80 & 124 & 11.3 & 2.30 & 1.09 & 2.71 & 7.11 \\
69 & 80 & 107 & 11.3 & 2.60 & 1.39 & 2.36 & 7.41 \\
70 & 140 & 153 & 11.2 & 2.45 & 0.80 & 2.90 & 7.30 \\
71 & 120 & 125 & 11.1 & 2.15 & 0.61 & 2.99 & 6.98 \\
72 & 140 & 128 & 11.1 & 2.40 & 0.75 & 2.85 & 7.25 \\
73 & 110 & 116 & 11.1 & 2.45 & 0.90 & 2.70 & 7.19 \\
74 & 100 & 79 & 10.9 & 1.00 & -0.46 & 3.86 & 5.75 \\
75 & 110 & 110 & 11.0 & 2.30 & 0.78 & 2.72 & 7.07 \\
76 & 130 & 145 & 11.3 & 1.00 & -0.54 & 4.29 & 5.90 \\
77 & 100 & 108 & 11.1 & 2.30 & 0.87 & 2.73 & 7.08 \\
78 & 100 & 64 & 10.9 & 2.70 & 1.23 & 2.17 & 7.44 \\
79 & 120 & 151 & 11.2 & 2.40 & 0.92 & 2.78 & 7.29 \\
80 & 100 & 123 & 11.1 & 1.20 & -0.24 & 3.84 & 5.97 \\
82 & 80 & 68 & 11.2 & 1.00 & -0.10 & 3.80 & 5.92 \\
83 & 80 & 79 & 11.1 & 2.30 & 1.20 & 2.40 & 7.22 \\
84 & 120 & 150 & 11.2 & 2.25 & 0.75 & 2.95 & 7.12 \\
86 & 80 & 84 & 11.1 & 1.00 & -0.10 & 3.70 & 5.92 \\
 \noalign{\smallskip} \hline \noalign{\smallskip} \\
 
\end{tabular}}
\end{table*}

\section*{Acknowledgements}
We thank the anonymous referee for a careful and professional review, for his/her insight into the nature of this object and its ULX relatives. This paper has been much improved through the referee's input. Dopita acknowledges the support from the Australian Department of Science and Education (DEST) Systemic Infrastructure Initiative grant and from an Australian Research Council (ARC) Large Equipment Infrastructure Fund (LIEF) grant LE0775546 which together made possible the construction of the WiFeS instrument. Dopita would also like to thank the Australian Research Council (ARC) for support under Discovery  project DP0984657. This research has made use of the NASA/IPAC Extragalactic Database (NED) which is operated by  the Jet Propulsion Laboratory, California Institute of Technology, under contract with the National  Aeronautics and Space Administration.  This research has also made use of NASA's Astrophysics Data System.

\end{document}